\documentclass[12pt,letterpaper]{JHEP3}
\usepackage{amssymb,amsmath,amsfonts,graphicx,graphics,eepic,epsfig}

\title{Casimir effect with non-local boundary conditions}
\author{Aram Saharian\\
Department of Physics, Yerevan State University,
1 Alex Manoogian Street, 375049 Yerevan, Armenia\\
Abdus Salam International Centre for Theoretical Physics,
34014 Trieste, Italy\\
E-mail: \email{saharian@ictp.it}}

\author{Giampiero Esposito\\
INFN, Sezione di Napoli, Complesso Universitario di Monte S.
Angelo, Via Cintia, Edificio N', 80126 Naples, Italy\\
Dipartimento di Scienze Fisiche, Complesso Universitario di
Monte S. Angelo, Via Cintia, Edificio N', 80126 Naples, Italy\\
E-mail: \email{giampiero.esposito@na.infn.it}}

\abstract{Non-local boundary conditions have been considered in
theoretical high-energy physics with emphasis on one-loop quantum
cosmology, one-loop conformal anomalies, Bose--Einstein
condensation models and spectral branes.
In the present paper, for the first time in
the literature, the Wightman function, the vacuum expectation
values of the field square and the energy-momentum tensor are
investigated for a massive scalar field satisfying non-local
boundary conditions on a single and two parallel plates. The
vacuum forces acting on the plates are evaluated. Interestingly,
suitable choices of the kernel in the non-local boundary
conditions lead to forces acting on the plates that can be
repulsive for intermediate distances. It is then possible to
obtain a locally stable equilibrium value of the interplate
distance stabilized by the vacuum forces.}


\begin{document}

\section{Introduction}

\label{sec:int}

In recent years, non-local boundary conditions in quantum physics
have been considered for at least three main purposes:
\vskip 0.3cm
\noindent
(i) As part of the attempt of obtaining a
consistent picture of one-loop quantum cosmology \cite{Espo99a,
Espo99b} and in the course of investigating one-loop conformal
anomalies for fermionic fields \cite{Deat91}.
\vskip 0.3cm
\noindent
(ii) Spectral boundary conditions for Laplace-type
operators on a compact manifold with boundary are partly
Dirichlet, partly (oblique) Neumann conditions, where the
partitioning is provided by a pseudodifferential projection; they
are of interest in string and brane theory \cite{Vasi01, Grub03}.
\vskip 0.3cm
\noindent
(iii) As part of the investigation of bulk
and surface states in Bose--Einstein condensation models
\cite{Schr89}.

In the latter case, following the work in Ref. \cite{Schr89}, it is useful
to consider a simple example given by the Laplacian acting on scalar
functions on the two-dimensional plane. More precisely, given the function $q
$ which is both Lebesgue summable and square-integrable on the real line,
i.e. $q\in L_{1}(\mathbf{R})\cap L_{2}(\mathbf{R})$, one defines \cite
{Schr89}
\begin{equation}
q_{R}(x)\equiv {\frac{1}{2\pi R}}\sum_{l=-\infty }^{\infty
}e^{ilx/R}\int_{-\infty }^{\infty }e^{-ily/R}q(y)dy.  \label{(1)}
\end{equation}
The function $q_{R}$ is, by construction, periodic with period $2\pi R$, and
tends to $q$ as $R$ tends to $\infty $. On considering the region
\begin{equation}
B_{R}\equiv \left\{ x,y:x^{2}+y^{2}\leq R^{2}\right\} ,  \label{(2)}
\end{equation}
one studies the Laplacian acting on square-integrable functions on $B_{R}$,
with non-local boundary conditions given by
\begin{equation}
\Bigr[u_{;N}\Bigr]_{\partial B_{R}}+\oint_{\partial B_{R}}q_{R}(s-s^{\prime
})u(R\cos (s^{\prime }/R),R\sin (s^{\prime }/R))ds^{\prime }=0.  \label{(3)}
\end{equation}
In polar coordinates, the resulting boundary-value problem reads as
\begin{equation}
-\left( {\frac{\partial ^{2}}{\partial r^{2}}}+{\frac{1}{r}}{\frac{\partial
}{\partial r}}+{\frac{1}{r^{2}}}{\frac{\partial ^{2}}{\partial \varphi ^{2}}}
\right) u=Eu,  \label{(4)}
\end{equation}
\begin{equation}
{\frac{\partial u}{\partial r}}(R,\varphi )+R\int_{-\pi }^{\pi
}q_{R}(R(\varphi -\theta ))u(R,\theta )d\theta =0.  \label{(5)}
\end{equation}
For example, when the eigenvalue $E$ is positive in Eq. (\ref{(4)}), the
corresponding eigenfunction is
\begin{equation}
u_{l,E}(r,\varphi )=J_{l}(r\sqrt{E})e^{il\varphi },  \label{(6)}
\end{equation}
where $J_{l}$ is the standard notation for the Bessel function of first kind
of order $l\in \mathbf{Z}$. On denoting by $\widetilde{q}$ the Fourier
transform of $q$, and inserting (\ref{(6)}) into the boundary condition (\ref
{(5)}), one finds an equation leading, \textit{implicitly}, to the knowledge
of the positive eigenvalues, i.e.
\begin{equation}
\Bigr[\sqrt{E}J_{l}^{\prime }(R\sqrt{E})+J_{l}(R\sqrt{E}){\widetilde{q}}(l/R)
\Bigr]=0.  \label{(7)}
\end{equation}
The solutions of Eq. (\ref{(7)}) which decay rapidly away from the boundary
are the surface states, whereas the solutions which remain non-vanishing are
called bulk states \cite{Schr89}.

In the extension to gauge fields, non-local boundary conditions along the
lines of Eq. (1.3) make it possible to improve the ellipticity properties
of the boundary-value problem, by working with suitable symbols of the
boundary operator, as shown in \cite{Espo00}. The price to be paid, however,
is that the gauge-field and ghost operators become pseudo-differential
because the gauge-fixing functional is no longer a local functional of
the gauge field \cite{Espo00}.

On the other hand, in the field-theoretical analysis of
macroscopic quantum effects such as the Casimir effect, an
essential point is the relation between the mode-sum energy,
evaluated as the sum of zero-point energies for each normal mode,
and the volume integral of the renormalized energy density. For
scalar fields with general curvature coupling it has been shown
that, in the discussion of this question for the Robin parallel
plates geometry it is necessary to include in the energy a surface
term concentrated on the boundary \cite{Rome02}. In subsequent
work, by using variational methods, the first author of the
present paper has derived an expression of the surface
energy-momentum tensor for a scalar field with a general curvature
coupling parameter in the general case of bulk and boundary
geometries \cite{Saha04}.

As a next step, we have been therefore led to consider the role of
non-local boundary conditions in the course of studying the vacuum
expectation value of the energy-momentum tensor as well as the
Casimir effect itself. For this purpose, section \ref{sec:1pl}
studies the Wightman function and Casimir densities for a single
plate, while section \ref{sec:2pl} is devoted to vacuum densities in
the region between two parallel plates. Concluding remarks are
made in section \ref{sec:conc}.

\section{Wightman function and Casimir densities for a single plate}

\label{sec:1pl}

We consider a real scalar field $\varphi (x)$ with general curvature
coupling parameter $\zeta $ satisfying the field equation
\begin{equation}
\left( \nabla _{\mu }\nabla ^{\mu }+m^{2}+\zeta R\right) \varphi =0,
\label{fieldeq}
\end{equation}
where $R$ is the scalar curvature for a $(D+1)$--dimensional
background spacetime, and $\nabla _{\mu }$ is the covariant
derivative operator. For special cases of minimally and
conformally coupled scalars one has $\zeta =0$ and $\zeta =\zeta
_c\equiv (D-1)/4D$, respectively. Our main interest in this paper
will be the Wightman function, the vacuum expectation values
(VEVs) of the field square and the energy-momentum tensor induced
by a single and two parallel plates in Minkowski spacetime. For
this problem the background spacetime is flat and in Eq.
(\ref{fieldeq}) we have $R=0$. As a result the eigenmodes are
independent of the curvature coupling parameter. However, the
local properties of the vacuum such as energy density and vacuum
stresses depend on this parameter.

In this section we consider the properties of the vacuum for the
geometry of a single plate. We will use rectangular coordinates
$x^{\mu }=(t,x^{1}=x, \mathbf{x}_{\parallel })$, where
$\mathbf{x}_{\parallel }=(x^{2},\ldots ,x^{D})$ denotes the
coordinates parallel to the plate. We assume that the plate is
located at $x=0$ and the field obeys a non-local boundary
condition similar to Eq. (\ref{(3)}), i.e.
\begin{equation}
n^{\nu }\partial _{\nu }\varphi (t,x,\mathbf{x}_{\parallel })
+\int d\mathbf{x}_{\parallel }^{\prime }
\,f(|\mathbf{x}_{\parallel }-\mathbf{x}_{\parallel
}^{\prime }|)\varphi (t,x,\mathbf{x}_{\parallel }^{\prime })=0,\;x=0,
\label{Boundcond}
\end{equation}
where $n^{\nu }$ is the inward-pointing normal to the boundary and
the conditions on the function $f$ will be specified below. For
definiteness we consider the region $x>0$ for which $n^{\nu
}=\delta _{1}^{\nu }$. It can be seen that, for this type of
boundary condition, the scalar product $(\varphi _{1},\varphi
_{2})_{t}$ for a given spatial hypersurface $t=\mathrm{const}$,
defined in the standard way (see, for instance, \cite{Birr82})
does not depend on the choice of hypersurface $\Sigma$.
Indeed, the corresponding difference for two hypersurfaces $
t=t_{1} $ and $t=t_{2}$, from the field equation, by using the
Stokes theorem, reads as
\begin{eqnarray}
\; & \; &
(\varphi _{1},\varphi _{2})_{t_{2}}-(\varphi _{1},\varphi
_{2})_{t_{1}} \nonumber \\
&=& i\int_{t_{1}}^{t_{2}}dt\int d\mathbf{x}_{\parallel }n^{\nu }
\left[ \varphi _{1}^{\ast }(t,0,\mathbf{x}_{\parallel })\partial _{\nu
}\varphi _{2}(t,0,\mathbf{x}_{\parallel })-\varphi _{2}(t,0,\mathbf{x}
_{\parallel })\partial _{\nu }\varphi _{1}^{\ast }(t,0,\mathbf{x}_{\parallel
})\right] .  \label{scproduct}
\end{eqnarray}
By virtue of the boundary condition (\ref{Boundcond}), the
integral on the right-hand side vanishes. For the standard local
Robin boundary conditions, the normal derivative of the field at a
given point on the boundary is determined by the value of the
field at the same point. The non-local boundary condition
(\ref{Boundcond}) states that the normal derivative at a given
point depends on the values of the field at other points on the
boundary. The properties of the boundary are codified by the
function $f$. In some sense, the situation here is similar to
that in electrodynamics for the spatial dispersion of the
dielectric permittivity, when one considers the relation between
the displacement and the electric field. In electrodynamics the
spatial dispersion leads to the dependence of dielectric
permittivity on the wave vector. Analogously, our non-local
boundary conditions lead to the dependence of the coefficient $F$
in the eigenfunctions on the wave vector ${\bf k}_{\parallel }$
(see below).

As the first stage in the investigation of local quantum effects
we consider the positive-frequency Wightman function. The
VEVs of the field square and the energy-momentum tensor can be
obtained from the Wightman function in the coincidence limit of
the arguments with an additional renormalization procedure.
Instead of the Wightman function we could take any other two-point
function, but we choose the Wightman function because it also determines
the response of particle detectors in a given state of motion. To
evaluate the positive-frequency Wightman function we use the
mode-sum formula
\begin{equation}
\langle 0_{S}|\varphi (x^{\mu })\varphi (x^{\prime \mu })|0_{S}\rangle
=\sum_{\mathbf{k}}\varphi _{\mathbf{k}}^{\ast }(x^{\mu })\varphi _{\mathbf{k}
}(x^{\prime \mu }),  \label{WFdef}
\end{equation}
where $|0_{S}\rangle $ is the vacuum state corresponding
to the geometry of a single plate. For this geometry, the normalised
eigenfunctions satisfying the boundary condition (\ref{Boundcond})
are given by
\begin{equation}
\varphi _{\mathbf{k}}(x^{\mu })=\frac{e^{i\mathbf{k}_{\parallel }\mathbf{x}
_{\parallel }-i\omega t}}{\sqrt{2^{D-1}\pi ^{D}\omega }}\cos (kx+\alpha ),
\label{eigfunc1}
\end{equation}
where $\omega \equiv \sqrt{k^{2}+k_{\parallel }^{2}+m^{2}}$,
$0\leq k<\infty $, \
and the function $\alpha =\alpha (k,k_{\parallel })$ is defined by the
relation
\begin{equation}
e^{2i\alpha } \equiv \frac{ik-F(k_{\parallel })}{ik+F(k_{\parallel })},
\label{alpha1}
\end{equation}
with $k_{\parallel }=|\mathbf{k}_{\parallel }|$. In the last formula we have
defined the Fourier transform
\begin{eqnarray}
F(k_{\parallel }) & \equiv &\int d\mathbf{x}_{\parallel}
\,f(|\mathbf{x}_{\parallel
}|)e^{i\mathbf{k}_{\parallel }\mathbf{x}_{\parallel }}  \notag \\
&=&\frac{(2\pi )^{\frac{D-1}{2}}}{k_{\parallel }^{\frac{D-3}{2}}}
\int_{0}^{\infty
}du\,u^{\frac{D-1}{2}}f(u)J_{\frac{D-3}{2}}(uk_{\parallel}),
\label{Fkpar}
\end{eqnarray}
where $J_{\nu }(x)$ is the Bessel function of first kind of order
$\nu$. In the case $F(k_{\parallel})>0$
there is also a purely imaginary eigenvalue $k=iF(k_{\parallel })$ with the
normalized eigenfunction
\begin{equation}
\varphi _{\mathbf{k}_{\parallel }}^{\mathrm{(im)}}(x^{\mu })=\sqrt{\frac{
F(k_{\parallel })}{(2\pi )^{D-1}\omega ^{\mathrm{(im)}}}}e^{i\mathbf{k}
_{\parallel }\mathbf{x}_{\parallel }-i\omega ^{\mathrm{(im)}
}t-xF(k_{\parallel })},  \label{imeigfunc}
\end{equation}
and $\omega ^{\mathrm{(im)}} \equiv \sqrt{k_{\parallel
}^{2}+m^{2}-F^{2}(k_{\parallel })}$, $k_{\parallel }^{2}\geq
F^{2}(k_{\parallel })-m^{2}$. These eigenfunctions correspond to the bound
states of the quantum field. The occurrence of purely imaginary eigenvalues
is proved by starting from the eigenfunctions, which depend on $x$ according
to $A_{1}e^{ikx}+A_{2}e^{-ikx}$, with constants $A_{i}$. From the boundary
condition one finds eventually
$$
A_{1}(ik+F(k_{||}))=A_{2}(ik-F(k_{||})).
$$
In addition to the solutions with both $A_{i} \not =0$, this equation has
indeed solutions $A_{1}=0, ik=F(k_{||})$ and
$A_{2}=0,ik=-F(k_{||})$.

To escape the instability of the vacuum state,
in the discussion below we will assume that the function (\ref{Fkpar})
satisfies the condition $F(k_{\parallel })\leq
\sqrt{k_{\parallel }^{2}+m^{2}}$.
For the convergence of the integral in (\ref{Fkpar}) we need to have the
behavior $f(u)=o(u^{1-D/2})$ in the limit $u\rightarrow \infty $ and the
behavior $f(u)=o(u^{1-D})$ in the limit $u\rightarrow 0$, by virtue
of standard summability criteria at infinity and at the origin, respectively.
For the discussion
below of various asymptotic cases it is useful to have the behavior of the
function $F(k_{\parallel })$ for large and small values of the argument. For
large values of $k_{\parallel }$,
by using the asymptotic expansion of
the Bessel function for large values of the argument, from (\ref{Fkpar}) we
see that $F(k_{\parallel })\sim o(k_{\parallel }^{\frac{3-D}{2}})$ as
$k_{\parallel }\rightarrow \infty $. For small values
of $k_{\parallel }$ two
subcases should be distinguished. For the case of functions $f(u)\sim
o(u^{1-D})$, $u\rightarrow \infty $, on using the expression for the Bessel
function for small values of the argument, one finds to leading order
\begin{equation}
F(k_{\parallel })\approx F_{0}\equiv \frac{2\pi ^{\frac{D-1}{2}}}{\Gamma
\left( \frac{D-1}{2}\right) }\int_{0}^{\infty
}du\,u^{D-2}f(u),\;k_{\parallel }\rightarrow 0.  \label{Fsmallt}
\end{equation}
For the functions $f(u)\sim u^{-\beta }$, $u\rightarrow \infty $,
with $ D/2-1<\beta <D-1$, by introducing in (\ref{Fkpar}) a new
integration variable $y=uk_{\parallel }$ and replacing the
function $f(y/k_{\parallel })$ by its asymptotic expansion for
large values of the argument, one finds $F(k_{||})\sim
k_{||}^{\beta +1-D}$, $k_{||} \rightarrow 0$. In the table below
we present examples of the kernel function $f$ in the boundary
condition (\ref{Boundcond}) with the corresponding Fourier
transforms $F$. These functions depend on three parameters
$f_{0}$, $x_{0}$, $\eta $.

\begin{table}
\caption{Examples of kernel function $f$ in the boundary condition, with
Fourier transform $F$.}
\begin{tabular}{|c|c|}
\hline
$f(x)$ & $F(y)$ \\ \hline
$\frac{f_{0}}{(x^{2}+x_{0}^{2})^{\frac{D-1}{2}-\eta }}$ & $f_{0}\frac{
2^{\eta +1}\pi ^{\frac{D-1}{2}}}{\Gamma \left( \frac{D-1}{2}-\eta \right) }
\left( \frac{x_{0}}{y}\right) ^{\eta }K_{\eta }(x_{0}y)$ \\ \hline
$f_{0}e^{-\eta \sqrt{x^{2}+x_{0}^{2}}}$ & $(2\pi x_{0})^{\frac{D}{2}}\eta
f_{0}\frac{K_{D/2}(x_{0}\sqrt{\eta ^{2}+y^{2}})}{\pi \left( \eta
^{2}+y^{2}\right) ^{D/4}}$ \\ \hline
$f_{0}\frac{e^{-\eta \sqrt{x^{2}+x_{0}^{2}}}}{\sqrt{x^{2}+x_{0}^{2}}}$ & $
2(2\pi x_{0})^{\frac{D}{2}-1}f_{0}\frac{K_{D/2-1}(x_{0}\sqrt{\eta ^{2}+y^{2}}
)}{\left( \eta ^{2}+y^{2}\right) ^{D/4-1/2}}$ \\ \hline
\end{tabular}
\end{table}

The positive-frequency Wightman function is obtained by substituting the
eigenfunctions (\ref{eigfunc1}) and (\ref{imeigfunc}) into the mode-sum
formula (\ref{WFdef}). It can be presented in the form
\begin{eqnarray}
\; & \; & \langle 0_{S}|\varphi (x^{\mu })
\varphi (x^{\prime \mu })|0_{S}\rangle
=\langle 0_{M}|\varphi (x^{\mu })\varphi (x^{\prime \mu })|0_{M}\rangle
\notag \\
&&+\int \frac{d\mathbf{k}_{\parallel }}{(2\pi )^{D}}e^{i\mathbf{k}
_{\parallel }(\mathbf{x}_{\parallel }-\mathbf{x}_{\parallel }^{\prime
})}\int_{0}^{\infty }dk\frac{e^{i\omega (t^{\prime }-t)}}{\omega }\cos \left[
k(x+x^{\prime })+2\alpha \right]  \notag \\
&&+\int \frac{d\mathbf{k}_{\parallel }}{(2\pi )^{D-1}}\frac{F(k_{\parallel })
}{\omega ^{\mathrm{(im)}}}\theta (F(k_{\parallel }))e^{i\mathbf{k}
_{\parallel }(\mathbf{x}_{\parallel }-\mathbf{x}_{\parallel }^{\prime
})-(x+x^{\prime })F(k_{\parallel })+i(t^{\prime }-t)\omega ^{\mathrm{(im)}}},
\label{WF1}
\end{eqnarray}
where $|0_{M}\rangle $ stands for the vacuum state in Minkowski
spacetime without boundary, with $\langle 0_{M}|\varphi (x^{\mu
})\varphi (x^{\prime \mu })|0_{M}\rangle $ the corresponding
Wightman function, and $\theta$ is the Heaviside step function.
The last term on the right of this formula is the contribution of
bound states. Note that, by using the definition for $\alpha
$, the $\cos $ function in the integrand can be also presented in
the form
\begin{equation}
\cos \left( ky+2\alpha \right) =\frac{k^{2}-F^{2}(k_{\parallel })}{
k^{2}+F^{2}(k_{\parallel })}\cos ky-\frac{2kF(k_{\parallel })}{
k^{2}+F^{2}(k_{\parallel })}\sin ky.  \label{rel1}
\end{equation}
The integrals in (\ref{WF1}) over the angular part of the vector
${\mathbf{k}}_{\parallel }$ can be explicitly evaluated with the
help of the formula already used in (\ref{Fkpar}). In the
coincidence limit from formula (\ref{WF1}) one finds the VEV of
the field square
\begin{eqnarray}
\langle 0_{S}|\varphi ^{2}|0_{S}\rangle &=&\langle 0_{M}|\varphi
^{2}|0_{M}\rangle +\frac{S_{D-1}}{(2\pi )^{D}}\int_{0}^{\infty
}dk_{\parallel }\,k_{\parallel }^{D-2}\int_{0}^{\infty }dk\frac{\cos \left(
2kx+2\alpha \right) }{\omega }  \notag \\
&&+\frac{S_{D-1}}{(2\pi )^{D-1}}\int dk_{\parallel }\,k_{\parallel }^{D-2}
\frac{F(k_{\parallel })}{\omega ^{\mathrm{(im)}}}\theta (F(k_{\parallel
}))e^{-2xF(k_{\parallel })},  \label{VEVphi2}
\end{eqnarray}
where $S_{D-1}=2\pi ^{(D-1)/2}/\Gamma ((D-1)/2)$ is the surface
area of the unit sphere in $D$-dimensional space. The last two
terms on the right of this formula are induced by the plate. For
points away from the boundary ($ x\neq 0$) these terms are finite
and the divergences in the VEV of the field square are contained
in the first term only. Hence, here the renormalization procedure
is the same as that for Minkowski spacetime without boundary.

To obtain an alternative form for the VEV of the field square, we write the
$\cos$ function in the second term on the right of (\ref{VEVphi2}) in terms
of the exponential functions and rotate the integration contour in the
complex $k$-plane by the angle $\pi /2$ for the term with $e^{2ikx}$ and by
the angle $-\pi /2$ for the term with $e^{-2ikx}$. We assume that the points
$\pm i\sqrt{k_{\parallel }^{2}+m^{2}}$ and the poles $\pm iF(k_{\parallel })$
in the case $F(k_{\parallel })>0$ are bypassed on the right by semicircles
with small radii. In such a way the following relation is obtained:
\begin{eqnarray}
\int_{0}^{\infty }dk\frac{\cos \left( 2kx+2\alpha \right) }{\omega }
&=&\int_{\sqrt{k_{\parallel }^{2}+m^{2}}}^{\infty }du\,\frac{e^{-2ux}}{\sqrt{
u^{2}-k_{\parallel }^{2}-m^{2}}}\frac{u+F(k_{\parallel })}{u-F(k_{\parallel
})}  \notag \\
&&-2\pi F(k_{\parallel })\frac{e^{-2xF(k_{\parallel })}\theta
(F(k_{\parallel }))}{\sqrt{k_{\parallel }^{2}+m^{2}-F^{2}(k_{\parallel })}}.
\label{intrel}
\end{eqnarray}
Now we see that the second term on the right of this formula cancels out the
third term on the right-hand side of formula (\ref{VEVphi2}) and we obtain
\begin{eqnarray}
\langle \varphi ^{2}\rangle _{\mathrm{sub}} &=&\langle 0_{S}|\varphi
^{2}|0_{S}\rangle -\langle 0_{M}|\varphi ^{2}|0_{M}\rangle =\frac{S_{D-1}}{
(2\pi )^{D}}\int_{0}^{\infty }dk_{\parallel }\,k_{\parallel }^{D-2}  \notag
\\
&&\times \int_{\sqrt{k_{\parallel }^{2}+m^{2}}}^{\infty }du
\,\frac{e^{-2ux}}{
\sqrt{u^{2}-k_{\parallel }^{2}-m^{2}}}\frac{u+F(k_{\parallel })}{
u-F(k_{\parallel })}.  \label{VEVphi21}
\end{eqnarray}
The VEVs of the field square in the cases of Dirichlet and Neumann boundary
conditions are obtained from the general formula (\ref{VEVphi21}) in the
limits $F(k_{\parallel })\rightarrow \infty $ and $F(k_{\parallel
})\rightarrow 0$, respectively. In these cases the integrals are evaluated
by introducing a new integration variable $v \equiv \sqrt{u^{2}-k_{\parallel
}^{2}-m^{2}}$ and passing to polar coordinates in the $(k_{\parallel },v)
$-plane. This simple calculation leads to the result
\begin{equation}
\left( \langle \varphi ^{2}\rangle _{\mathrm{sub}}\right) _{\mathrm{Dirichlet
}}=-\left( \langle \varphi ^{2}\rangle _{\mathrm{sub}}\right) _{\mathrm{
Neumann}}=-\frac{(m/x)^{\frac{D-1}{2}}}{2^{D}\pi ^{\frac{D+1}{2}}}
K_{(D-1)/2}(2mx),  \label{phi2Dir}
\end{equation}
where $K_{\nu }(x)$ is the modified Bessel function of second
kind. The VEV of the field square induced by a single plate on the
background spacetime $ R^{(D,1)}\times \Sigma $, with an internal
space $\Sigma $ and local Robin boundary condition, is
investigated in \cite{Saha05a} as a limiting case of the
braneworld geometry.

Having the Wightman function (\ref{WF1}) and the VEV of the field
square we can evaluate the VEV\ of the energy-momentum tensor by
making use of the formula
\begin{equation}
\langle 0_{S}|T_{\mu \nu }|0_{S}\rangle =\lim_{x^{\prime }\rightarrow
x}\langle 0_{S}|\partial _{\mu }\varphi (x^{\alpha })\partial _{\nu
}^{\prime }\varphi (x^{\prime \alpha })|0_{S}\rangle +\left[ \left( \zeta -
\frac{1}{4}\right) g_{\mu \nu }\partial _{\alpha }\partial ^{\alpha }-\zeta
\partial _{\mu }\partial _{\nu }\right] \langle 0_{S}|\varphi
^{2}|0_{S}\rangle .  \label{VEVemt}
\end{equation}
With the help of formulae (\ref{WF1}) and (\ref{VEVphi21}) for the boundary
induced part the following result is obtained (no summation over $\nu $):
\begin{eqnarray}
\langle T_{\mu }^{\nu }\rangle _{\mathrm{sub}} &=&\langle 0_{S}|T_{\mu
}^{\nu }|0_{S}\rangle -\langle 0_{M}|T_{\mu }^{\nu }|0_{M}\rangle =\frac{
S_{D-1}\delta _{\mu }^{\nu }}{(2\pi )^{D}}\int_{0}^{\infty }dk_{\parallel
}\,k_{\parallel }^{D-2}  \notag \\
&&\times \int_{\sqrt{k_{\parallel }^{2}+m^{2}}}^{\infty }du\,\frac{A_{\nu
}(u,k_{\parallel })e^{-2ux}}{\sqrt{u^{2}-k_{\parallel }^{2}-m^{2}}}\frac{
u+F(k_{\parallel })}{u-F(k_{\parallel })},  \label{VEVemt1}
\end{eqnarray}
where we have introduced the notations
\begin{eqnarray}
A_{0}(u,k_{\parallel }) &\equiv &k_{\parallel }^{2}+m^{2}-4\zeta
u^{2},\;A_{1}(u,k_{\parallel })=0,  \label{A0} \\
A_{\nu }(u,k_{\parallel }) &\equiv &(1-4\zeta )u^{2}-\frac{k_{\parallel }^{2}
}{(D-1)},\;\nu =2,\ldots ,D.  \label{Anu}
\end{eqnarray}
For the evaluation of the $T_{\nu}^{\nu}$ components, with $\nu=2,3,...$,
we note that, from the problem symmetry, one has $T_{2}^{2}=T_{3}^{3}=...$
and hence
$$
T_{2}^{2}=\sum_{\nu=2}^{D}{T_{\nu}^{\nu}\over (D-1)}.
$$
For the sum in the last formula, the integrand contains the factor
$$
\sum_{\nu=2}^{D}T_{\nu}^{\nu} \rightarrow
-k_{||}^{2}+u^{2}(1-4 \zeta)(D-1).
$$
As we see, the vacuum stress in the direction perpendicular to the
plate, $\langle T_{1}^{1}\rangle _{\mathrm{sub}}$, vanishes. This
result could be also directly obtained from the continuity
equation $\partial _{\nu }\langle T_{\mu }^{\nu }\rangle _{
\mathrm{sub}}=0$ for the boundary induced VEV of the
energy-momentum tensor. It can be checked that the tensor
(\ref{VEVemt1}) is traceless for a conformally coupled massless
scalar field. In the case of the local boundary condition of Robin
type the function $F(k_{\parallel })$ is a constant and the
expression for the corresponding energy-momentum tensor is further
simplified by introducing a new integration variable $v \equiv
\sqrt{ u^{2}-k_{\parallel }^{2}-m^{2}}$ and passing to polar
coordinates in the $(k_{\parallel },v)$-plane
as before (2.15). In particular, for
the massless case we obtain the result given in \cite{Rome02}.
Note that in this case the VEV of the energy-momentum tensor
vanishes for a conformally coupled scalar field. For the non-local
boundary condition, in general, this is not the case. Another
important difference is that, unlike the local case, for non-local
boundary conditions, in general, $\langle T_{0}^{0}\rangle
_{\mathrm{sub} }\neq \langle T_{\mu }^{\mu }\rangle
_{\mathrm{sub}}$ (no summation over $ \mu =2,\ldots ,D$). For
Dirichlet and Neumann boundary conditions the vacuum
energy-momentum tensor is further simplified by a method similar to
that used for the VEV of the field square, and one finds (no
summation over $\mu $)
\begin{eqnarray}
\; & \; &
\left( \langle T_{\mu }^{\mu }\rangle _{\mathrm{sub}}\right) _{\mathrm{
Dirichlet}} =-\left( \langle T_{\mu }^{\mu }\rangle _{\mathrm{sub}}\right)
_{\mathrm{Neumann}}  \notag \\
&=&\frac{(m/x)^{\frac{D+1}{2}}}{2^{D}\pi ^{\frac{D+1}{2}}}\left[ 2D(\zeta
-\zeta _{c})K_{\frac{D+1}{2}}(2mx)+(4\zeta -1)mxK_{\frac{D-1}{2}}(2mx)\right]
,  \label{EMTDir}
\end{eqnarray}
$\mu =0,2,\ldots ,D$. In the Dirichlet case the corresponding
energy density is negative everywhere for both minimally and
conformally coupled scalars. The vacuum energy-momentum tensor for
a plate on the background spacetime $ R^{(D,1)}\times \Sigma $,
with an internal space $\Sigma$ and local Robin boundary condition
is investigated in \cite{Saha05b}.

Now let us consider the limiting cases of the VEVs
(\ref{VEVphi21}), (\ref {VEVemt1}). These VEVs diverge on the
boundary. Surface divergences are well-known in quantum field
theory with boundaries and are investigated for various types of
boundary geometries and boundary conditions 
(see, for instance, \cite
{Birr82,Deut79}). They result from the idealization of the
boundaries as perfectly smooth surfaces which are perfect
reflectors at all frequencies. It seems plausible that such
effects as surface roughness, or the microstructure of the
boundary on small scales can introduce a physical cutoff needed to
produce finite values of surface quantities. An alternative
mechanism for introducing a cutoff which removes singular behaviour
on boundaries is to allow the position of the boundary to undergo
quantum fluctuations \cite{Ford98}. Such fluctuations smear out
the contribution of the high-frequency modes without the need to
introduce an explicit high-frequency cutoff. Note that in this
paper we consider boundary induced vacuum densities which are
finite away from the boundary. We expect that similar results
would be obtained in the model where instead of externally imposed
boundary condition the fluctuating field is coupled to a smooth
background potential that implements the boundary condition in a
certain limit \cite{Grah02}. In the problem under consideration,
for the points near the plate, the main contribution results from
large values of $u$. To leading order we can omit $F(k_{\parallel
})$ in the integrands and the VEVs behave as those for the Neumann
boundary conditions, i.e.
\begin{equation}
\langle \varphi ^{2}\rangle _{\mathrm{sub}}\approx \frac{x^{1-D}}{\left(
4\pi \right) ^{\frac{D+1}{2}}}\Gamma \left( \frac{D-1}{2}\right) ,\quad
\langle T_{\mu }^{\mu }\rangle _{\mathrm{sub}}\approx \frac{2D(\zeta
_{c}-\zeta )}{\left( 4\pi \right) ^{\frac{D+1}{2}}x^{D+1}}\Gamma \left(
\frac{D+1}{2}\right) .  \label{near1plate}
\end{equation}
For large distances from the plate the main contribution to the VEVs comes
from small values of $u$. It can be seen that, for a massless field or for a
massive field with $f(u)\sim u^{-\beta }$, $D/2-1<\beta <D-1$, at large
distances the VEVs behave as those for Dirichlet boundary condition. For the
case of massive field with $f(u)\sim o(u^{1-D})$, to leading order
one has
\begin{equation}
\langle \varphi ^{2}\rangle _{\mathrm{sub}}\approx \frac{F_{0}+m}{F_{0}-m}
\left( \langle \varphi ^{2}\rangle _{\mathrm{sub}}\right) _{\mathrm{
Dirichlet }},\;x\rightarrow \infty ,  \label{phi2largedist}
\end{equation}
and a similar relation for the VEV of the energy-momentum tensor. Here
$F_{0}$ is defined by relation (\ref{Fsmallt}).

We have done numerical calculations for the components of the vacuum
energy-momentum tensor by taking, for example, the kernel function
\begin{equation}
f(x) \equiv f_{0}e^{-\eta x}.  \label{fxexamp}
\end{equation}
The corresponding Fourier transform $F(k_{\parallel })$
is obtained from the second
line of the table given earlier in the limit $x_{0}\rightarrow 0$, i.e.
\begin{equation}
F(k_{\parallel })=\frac{\eta F_{1}}{\left( 1+k_{\parallel
}^{2}/\eta ^{2}\right) ^{D/2}},  \label{Fkexamp}
\end{equation}
with the notation $F_{1} \equiv 2^{D-1}\pi ^{\frac{D}{2}-1}
\Gamma (D/2)f_{0}/\eta^{D}$.
In figure 1 we have presented the vacuum energy density
(full curves) and ${T}_{2}^{2}$-stress (dashed curves) as functions on $\eta
x $ for various values of the parameter $F_{1}$ (numbers near the curves)
for a minimally coupled $D=3$ scalar field. The energy density is positive
for the region near the plate and is negative at large distances from the
plate, having the negative minimum for some intermediate distance. The
corresponding curves for the energy density of the conformally coupled
scalar field are presented in figure 2. In this case the
${T}_{2}^{2}$-stress is related to the energy density by the traceless
condition: $\langle T_{2}^{2}\rangle _{\mathrm{sub}}=-\langle
T_{0}^{0}\rangle _{\mathrm{sub}}/2$. Note that, for the case of a conformally
coupled massless scalar field with local boundary condition, the
corresponding vacuum energy-momentum tensor vanishes.

\EPSFIGURE[htp]{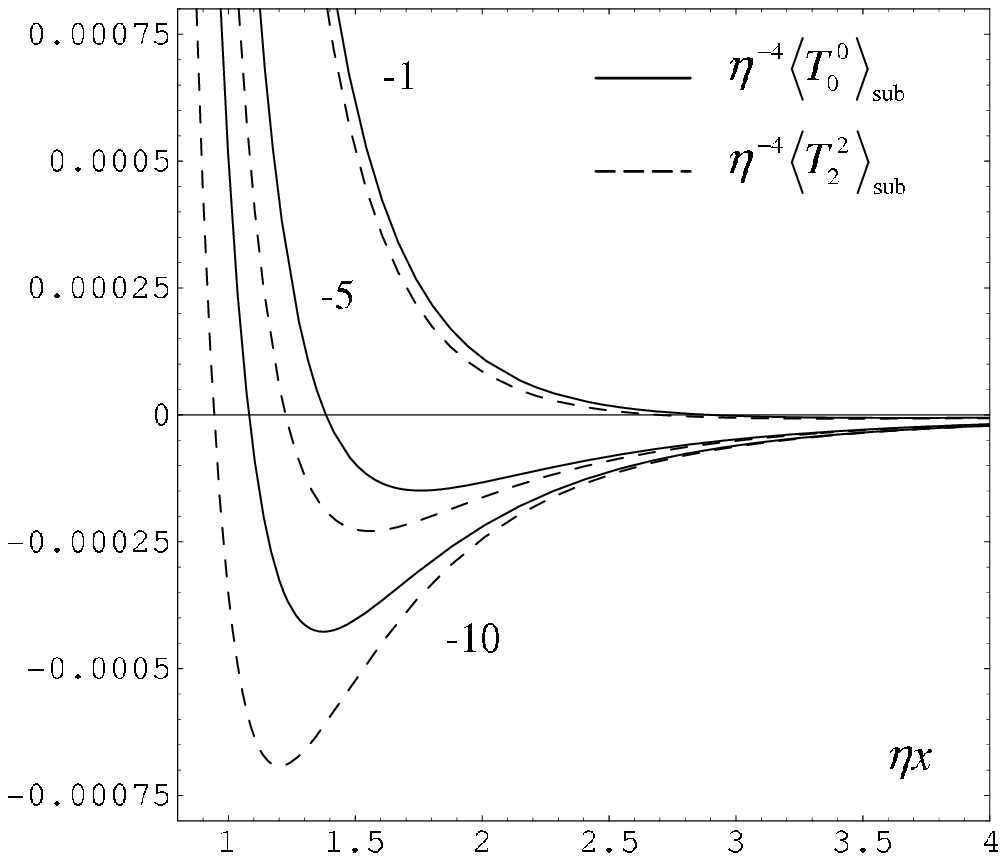,width=.6\textwidth}{Single-plate
induced vacuum densities $\langle T_{0}^{0}
\rangle_{\mathrm{sub}}/\protect\eta^{D+1}$ (full curves) and
$\langle T_{2}^{2}\rangle_{\mathrm{sub}}/\protect\eta^{D+1}$
(dashed curves) as functions of $\protect\eta x$ for a minimally
coupled massless scalar in $D=3$ in the case of kernel function
(\protect\ref{fxexamp}). The numbers near the curves are the
values of the parameter $F_{1}$ defined in the paragraph after
formula (\protect\ref{Fkexamp}).}

\EPSFIGURE[htp]{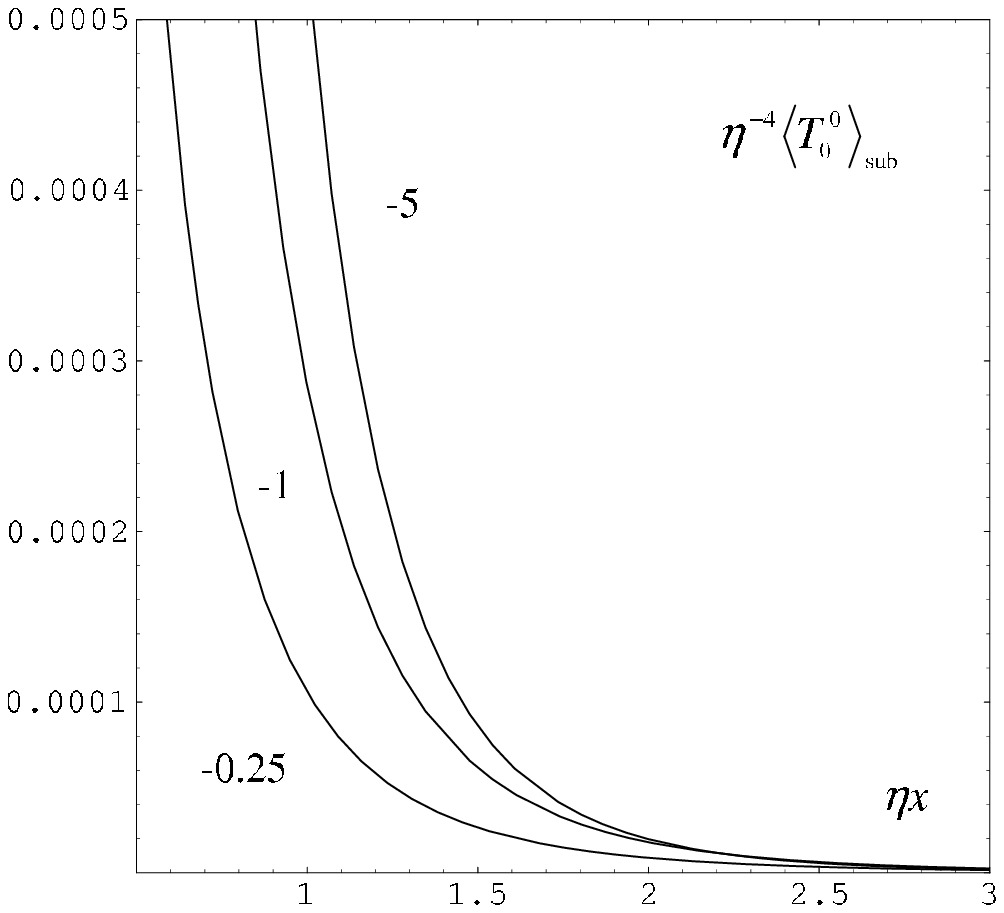,width=.6\textwidth}{Single-plate
induced vacuum energy density $\langle T_{0}^{0}
\rangle_{\mathrm{sub}}/\protect\eta^{D+1}$ as a function of
$\protect\eta x$ for a conformally coupled massless scalar in
$D=3$ in the case of kernel function (\protect\ref{fxexamp}). The numbers
near the curves are the values of the parameter $F_{1}$ defined in
the paragraph after formula (\protect\ref{Fkexamp}). The vacuum
stress $\langle T_{2}^{2}\rangle_{\mathrm{sub}}$ is expressed
through the energy density by the traceless condition.}

\clearpage

\section{Vacuum densities in the region between two parallel plates}

\label{sec:2pl}

\subsection{Eigenfunctions}

In this section we investigate the VEVs for the geometry of two parallel
plates. We assume that the plates are located at $x=a_{j}$, $j=1,2$ and
the field obeys non-local boundary conditions
\begin{equation}
n_{(j)}^{\nu }\partial _{\nu }\varphi (x^{\mu })+\int d\mathbf{x}_{\parallel
}^{\prime }\,f_{j}(|\mathbf{x}_{\parallel }-\mathbf{x}_{\parallel }^{\prime
}|)\varphi (x^{\prime \mu })=0,\;x=a_{j},  \label{Boundcond2}
\end{equation}
where $n_{(j)}^{\nu }$ is the inward-pointing unit normal to the boundary
at $x=a_{j}$. Here we consider the general case when the kernel functions
$f_{j}$ determining the properties of the boundaries are
different for the plates. The VEVs in the regions $x<a_{1}$ and $x>a_{2}$
coincide with the corresponding quantities for a single plate geometry and
are investigated in the previous section. Here we consider the region
between the plates with $n_{(j)}^{\nu }=(-1)^{j-1}\delta _{1}^{\nu }$. In
this region the corresponding eigenfunctions are presented in two equivalent
forms corresponding to $j=1,2$, i.e.
\begin{equation}
\varphi _{\mathbf{k}}(x^{\mu })=\beta (k,k_{\parallel })e^{i\mathbf{k}
_{\parallel }\mathbf{x}_{\parallel }-i\omega t}\cos \left[ k|x-a_{j}|+\alpha
_{j}\right] ,  \label{eigfunc2pl}
\end{equation}
where $\alpha _{j}=\alpha _{j}(k,k_{\parallel })$ is defined by the relation
\begin{equation}
e^{2i\alpha _{j}(k,k_{\parallel })}\equiv \frac{ik-(-1)^{j-1}F_{j}(k_{
\parallel })}{ik+(-1)^{j-1}F_{j}(k_{\parallel })},  \label{alfj}
\end{equation}
with the definition of Fourier transform $F_{j}(k_{||})$
similar to Eq. (\ref{Fkpar}), i.e.
\begin{eqnarray}
F_{j}(k_{\parallel }) & \equiv &\int d\mathbf{x}_{\parallel }
\,f_{j}(|\mathbf{x}%
_{\parallel }|)e^{i\mathbf{k}_{\parallel }\mathbf{x}_{\parallel }}  \notag \\
&=&\frac{(2\pi )^{\frac{D-1}{2}}}{k_{\parallel }^{\frac{D-3}{2}}}
\int_{0}^{\infty
}du\,u^{\frac{D-1}{2}}f_{j}(u)J_{\frac{D-3}{2}}(uk_{\parallel }).
\label{Fj}
\end{eqnarray}
The corresponding eigenvalues $z=ka$, $a=a_{2}-a_{1}$, are solutions of the
following transcendental equation:
\begin{equation}
\left( z^{2}-c_{1}c_{2}\right) \sin z+\left( c_{1}+c_{2}\right) z\cos z=0,
\label{eigvalues}
\end{equation}
where the coefficients $c_{j}$ are defined by the relations
\begin{equation}
c_{j}\equiv (-1)^{j-1}aF_{j}(k_{\parallel }).  \label{cj}
\end{equation}
Equation (\ref{eigvalues}) is obtained by bearing in mind that the
$x$-dependence in the eigenfunctions has the
form $A_{1}e^{ikx}+A_{2}e^{-ikx}$.
The boundary conditions (\ref{Boundcond2}) lead therefore to the linear
homogeneous system
\begin{equation}
(F_{1}+ik)e^{ika_{1}}A_{1}+(F_{1}-ik)e^{-ika_{1}}A_{2}=0,
\end{equation}
\begin{equation}
(F_{2}+ik)e^{ika_{2}}A_{1}+(F_{2}-ik)e^{-ika_{2}}A_{2}=0.
\end{equation}
Non-trivial solutions for $A_{1}$ and $A_{2}$ exist if and only if Eq. (\ref
{eigvalues}) holds. The expression for the coefficient $\beta
(k,k_{\parallel })$ in (\ref{eigfunc2pl}) is obtained from the normalization
condition:
\begin{equation}
\beta ^{-2}(ka,k_{\parallel })=(2\pi )^{D-1}a\omega \left[ 1+\frac{\sin (ka)
}{ka}\cos (ka+2\alpha _{j})\right] .  \label{bet2pl}
\end{equation}
The eigenvalue condition (\ref{eigvalues}) has an infinite set of real zeros
which we will denote by $z=\lambda _{n}$, $n=1,2,\ldots $. In addition,
depending on the values of the coefficients $c_{j}$, this equation has two
or four complex conjugate purely imaginary zeros (see, for instance, \cite
{Rome02}) $\pm iy_{l}$, $y_{l}>0$.

\subsection{Wightman function}

Substituting the eigenfunctions (\ref{eigfunc2pl}) into the corresponding
mode-sum formula, for the positive-frequency Wightman function in the region
between two plates one finds
\begin{equation}
\langle 0|\varphi (x^{\mu })\varphi (x^{\prime \mu })|0\rangle =\int d
\mathbf{k}_{\parallel }e^{i\mathbf{k}_{\parallel }(\mathbf{x}_{\parallel }-
\mathbf{x}_{\parallel }^{\prime })}\sum_{z=\lambda _{n},iy_{l}}\frac{
e^{i\omega (t^{\prime }-t)}}{\beta ^{2}(z,k_{\parallel })}\cos
(zx_{j}+\alpha _{j})\cos (zx_{j}^{\prime }+\alpha _{j}),  \label{WF2pl}
\end{equation}
where $|0\rangle$ is the vacuum state in the region
between the plates and we use the notations
\begin{equation}
x_{j}\equiv \frac{|x-a_{j}|}{a},\;x_{j}^{\prime }\equiv \frac{|x^{\prime
}-a_{j}|}{a}.  \label{xj}
\end{equation}
The summation over the eigenvalues $\lambda _{n}$, $iy_{l}$ can be done by
using the formula
\begin{eqnarray}
\sum_{z=\lambda _{n},iy_{l}}\frac{h(z)}{1+\cos (z+2\alpha _{1})\sin z/z} &=&-
\frac{1}{2}\frac{h(0)}{1-c_{1}^{-1}-c_{2}^{-1}}+\frac{1}{\pi }
\int_{0}^{\infty }dzh(z)  \notag \\
&&+\frac{i}{\pi }\int_{0}^{\infty }dt\frac{h(te^{\pi i/2})-h(te^{-\pi i/2})}{
\frac{(t-c_{1})(t-c_{2})}{(t+c_{1})(t+c_{2})}e^{2t}-1}  \notag \\
&&-\frac{\pi \theta (c_{j})}{2c_{j}}\left[ g_{j}(c_{j}e^{\pi
i/2})+g_{j}(c_{j}e^{-\pi i/2})\right] ,  \label{sumformula}
\end{eqnarray}
where
\begin{equation}
g_{j}(z)\equiv \left( z^{2}+c_{j}^{2}\right) h(z).  \label{gjz}
\end{equation}
Formula (\ref{sumformula}) is derived in \cite{Rome02} as a special case of
the generalized Abel--Plana summation formula \cite{SahaAP}. As a function
$h$ (with first-order poles at $z=\pm ic_{j}$) we take
\begin{equation}
h(z)\equiv \frac{e^{i\omega (t^{\prime }-t)}}{a\omega }\cos (zx_{j}+\alpha
_{j})\cos (zx_{j}^{\prime }+\alpha _{j}),\;
\omega \equiv \sqrt{z^{2}/a^{2}+k_{
\parallel }^{2}+m^{2}},  \label{hz}
\end{equation}
where the function $\alpha _{j}=\alpha _{j}(z/a,k_{\parallel })$ is defined
by formula (\ref{alfj}). By using the relation (from Eq. (\ref{rel1}))
\begin{equation}
\cos (y+2\alpha _{j})=\frac{z^{2}-c_{j}^{2}}{z^{2}+c_{j}^{2}}\cos y-\frac{
2zc_{j}}{z^{2}+c_{j}^{2}}\sin y,  \label{cos1}
\end{equation}
it can be seen that
\begin{equation}
g_{j}(c_{j}e^{\pi i/2})+g_{j}(c_{j}e^{-\pi i/2})=2c_{j}^{2}\frac{e^{i\omega
^{\mathrm{(im)}}(t^{\prime }-t)}}{a\omega ^{\mathrm{(im)}}}
e^{-c_{j}(x_{j}+x_{j}^{\prime })}.  \label{gjcj}
\end{equation}
Moreover, by making use of the definition for $\alpha _{j}$ we see that
$e^{2i\alpha _{j}(0,k_{\parallel })}=-1$, and hence $\cos (2\alpha
_{j}(0,k_{\parallel }))=-1$. This implies in turn that $h(0)=0$. The
resulting Wightman function from (\ref{WF2pl}) is found to be
\begin{eqnarray}
\langle 0|\varphi (x^{\mu })\varphi (x^{\prime \mu })|0\rangle &=&\langle
0_{S}|\varphi (x^{\mu })\varphi (x^{\prime \mu })|0_{S}
\rangle _{j}+\frac{4}{
(2\pi )^{D}}\int d\mathbf{k}_{\parallel }
e^{i\mathbf{k}_{\parallel }(\mathbf{
x}_{\parallel }-\mathbf{x}_{\parallel }^{\prime })}  \notag \\
&&\times \int_{a\sqrt{k_{\parallel }^{2}+m^{2}}}^{\infty }dt\frac{\cosh
(tx_{j}+\tilde{\alpha}_{j})\cosh (tx_{j}^{\prime }+\tilde{\alpha}_{j})}{
\frac{(t-c_{1})(t-c_{2})}{(t+c_{1})(t+c_{2})}e^{2t}-1}  \notag \\
&&\times \frac{\cosh \left[ (t-t^{\prime })\sqrt{t^{2}/a^{2}-k_{\parallel
}^{2}-m^{2}}\right] }{\sqrt{t^{2}-k_{\parallel }^{2}a^{2}-m^{2}a^{2}}},
\label{WF2pl1}
\end{eqnarray}
where the function $\tilde{\alpha}_{j}=\tilde{\alpha}_{j}(t,k_{\parallel })$
is defined by the relation
\begin{equation}
e^{2\tilde{\alpha}_{j}}\equiv \frac{t-c_{j}}{t+c_{j}},  \label{alfajtilde}
\end{equation}
and $\langle 0_{S}|\varphi (x^{\mu })\varphi (x^{\prime \mu })|0_{S}\rangle
_{j}$ is the Wightman function for a single plate located at $x=a_{j}$.
On taking the coincidence limit, for the VEV of the field square we obtain the
formula
\begin{eqnarray}
\langle 0|\varphi ^{2}|0\rangle &=&\langle 0_{S}|\varphi ^{2}|0_{S}\rangle
_{j}+\frac{4S_{D-1}}{(2\pi )^{D}}\int_{0}^{\infty }dk_{\parallel
}k_{\parallel }^{D-2}\int_{a\sqrt{k_{\parallel }^{2}+m^{2}}}^{\infty }dt
\frac{\cosh ^{2}(tx_{j}+\tilde{\alpha}_{j})}{\sqrt{t^{2}-k_{\parallel
}^{2}a^{2}-m^{2}a^{2}}}  \notag \\
&&\times \left[ \frac{(t-c_{1})(t-c_{2})}{(t+c_{1})(t+c_{2})}e^{2t}-1\right]
^{-1},  \label{phi22pl}
\end{eqnarray}
where $\langle 0_{S}|\varphi ^{2}|0_{S}\rangle _{j}$ is the corresponding
VEV for the geometry of a single plate at $x=a_{j}$ and is investigated in
the previous section. The surface divergences
on the plate at $x=a_{j}$ are
contained in this term. The second term on the right of formula (\ref
{phi22pl}) is finite at $x=a_{j}$ and is induced by the second plate located
at $x=a_{j_{1}}$, $j_{1}=1,2$, $j_{1}\neq j$. This term diverges at
$x=a_{j_{1}}$. The corresponding divergence is the same as that for the
geometry of a single plate located at $x=a_{j_{1}}$.

\subsection{VEV of the energy-momentum tensor and vacuum forces}

The vacuum expectation value of the energy-momentum tensor is evaluated by
formula (\ref{VEVemt}) with the vacuum state $|0\rangle$. By taking into
account formulae (\ref{WF2pl1}), (\ref{phi22pl}), for the region between the
plates one finds (no summation over $\nu$)
\begin{eqnarray}
\langle 0|T_{\mu }^{\nu }|0\rangle  &=&\langle 0_{S}|T_{\mu }^{\nu
}|0_{S}\rangle _{j}+\delta _{\mu }^{\nu }\frac{2S_{D-1}}{(2\pi )^{D}}
\int_{0}^{\infty }dk_{\parallel }k_{\parallel }^{D-2}\int_{a\sqrt{
k_{\parallel }^{2}+m^{2}}}^{\infty }\frac{f_{j\nu }(t,k_{\parallel },x)dt}{
\sqrt{t^{2}-k_{\parallel }^{2}a^{2}-m^{2}a^{2}}}  \notag \\
&&\times \left[ \frac{(t-c_{1})(t-c_{2})}{(t+c_{1})(t+c_{2})}e^{2t}-1\right]
^{-1},  \label{Tmunu2pl}
\end{eqnarray}
where $\langle 0_{S}|T_{\mu }^{\nu }|0_{S}\rangle _{j}$ is the vacuum
energy-momentum tensor for the geometry of a
single plate located at $x=a_{j}$,
and the second term on the right is the part of the energy-momentum tensor
induced by the presence of the second plate. In formula (\ref{Tmunu2pl}) we
have defined
\begin{eqnarray}
f_{j0}(t,k_{\parallel },x) &\equiv &k_{\parallel
}^{2}+m^{2}-t^{2}/a^{2}+\left( k_{\parallel }^{2}+m^{2}-4\zeta
t^{2}/a^{2}\right) \cosh \left( 2tx_{j}+2\tilde{\alpha}_{j}\right) ,
\label{fj0} \\
f_{j1}(t,k_{\parallel },x) &\equiv &t^{2}/a^{2},  \label{fj1} \\
f_{j\nu }(t,k_{\parallel },x) &\equiv &
-\frac{k_{\parallel }^{2}}{(D-1)}-\left[
\frac{k_{\parallel }^{2}}{(D-1)}+(4\zeta -1)\frac{t^{2}}{a^{2}}\right] \cosh
\left( 2tx_{j}+2\tilde{\alpha}_{j}\right) ,  \label{fjnu}
\end{eqnarray}
with $\nu =2,3,\ldots ,D$. It can be easily checked that the vacuum
energy-momentum tensor is traceless for a conformally coupled massless
scalar field. As we could expect from the problem symmetry, the vacuum
stresses in the directions parallel to the plates are isotropic. Note that
the $T_{1}^{1}$-component of the energy-momentum tensor is uniform. This
also follows from the continuity equation for the VEV of the energy-momentum
tensor. In particular, the $T_{1}^{1}$-stress is finite on the plates and
determines the vacuum force acting per unit surface of the plate:
\begin{eqnarray}
p &=&-\langle 0|T_{1}^{1}|0\rangle =-\frac{2S_{D-1}}{(2\pi )^{D}}
\int_{0}^{\infty }du\,u^{D-2}\int_{\sqrt{u^{2}+m^{2}}}^{\infty }\frac{t^{2}dt
}{\sqrt{t^{2}-u^{2}-m^{2}}}  \notag \\
&&\times \left[ \frac{(t-F_{1}(u))(t+F_{2}(u))}{(t+F_{1}(u))(t-F_{2}(u))}
e^{2at}-1\right] ^{-1}.  \label{vacforce}
\end{eqnarray}
When the functions $F_{j}(u)=\mathrm{const}$, this formula can be
simplified by introducing a new integration variable $v \equiv \sqrt{
t^{2}-u^{2}-m^{2}}$ and passing to polar coordinates in the $(u,v)$ plane.
After integrating over the angular part one finds the formula
\begin{equation}
p=-\frac{2^{1-D}\pi ^{-D/2}}{\Gamma (D/2)}\int_{m}^{\infty }dt\frac{
t^{2}\left( t^{2}-m^{2}\right) ^{\frac{D}{2}-1}}{\frac{(t-F_{1})(t+F_{2})}{
(t+F_{1})(t-F_{2})}e^{2at}-1}.  \label{vacforceFconst}
\end{equation}
For a massless scalar field this result coincides with the formula
obtained in \cite{Rome02}. The VEVs of the field square and the
energy-momentum tensor for the geometry of two parallel plates on
the background spacetime $R^{(D_{1},1)}\times \Sigma$ with an
internal space $\Sigma $ and local Robin boundary conditions are
investigated in \cite{Saha05a,Saha05b}. The vacuum forces for
Dirichlet and Neumann boundary conditions are the same and are
obtained from (\ref{vacforceFconst}) as special cases with
$F_{j}=\infty $ and $F_{j}=0$, respectively. These forces are
attractive for all interplate distances.

Now we turn to the discussion of the general formula
(\ref{vacforce}) for the vacuum forces in the limiting cases
corresponding to small and large interplate distances. For this it
is convenient to introduce a new integration variable $v \equiv
\sqrt{t^{2}-u^{2}-m^{2}}$ as above and to pass to polar
coordinates $(r,\theta )$ in the $(u,v)$ plane. For small
distances the main contribution to the $r$-integral comes from
large values of $r$. From the asymptotic behavior of the functions
$F_{j}(u)$ for large values of the argument described in section
\ref{sec:1pl}, the functions $F_{j}(u)$ in the coefficient of
$e^{2at}$ may be omitted and, to leading order, the corresponding
forces coincide with the vacuum forces in the case of Neumann
boundary conditions and are attractive. For large distance between
the plates the main contribution to the integral results from
small values of $r$ and two subcases should be distinguished. For a
massless scalar field or for a massive field with kernel functions
$f_{j}(u)\sim u^{-\beta _{j}}$, $D/2-1<\beta _{j}<D-1$,
$u\rightarrow \infty$, the terms with $t$ in the coefficient of
$e^{2at}$ may be omitted and the vacuum forces coincide with those
for the Dirichlet case and are attractive. In the second subcase,
corresponding to the kernel functions $f_{j}(u)\sim o(u^{1-D})$,
$u\rightarrow \infty $, to leading order the vacuum forces are
given by expressions which are obtained from (\ref{vacforce}) by
the replacements $t\pm F_{j}(u)\rightarrow m\pm F_{j0}$ in the
coefficient of $ e^{2at}$, where the constants $F_{j0}$ are
defined by formulae similar to (\ref{Fsmallt}) with the
replacements $f(u)\rightarrow f_{j}(u)$.

We have evaluated numerically the vacuum forces acting on the plates in the
case of the kernel functions
\begin{equation}
f_{j}(x) \equiv f_{0j}e^{-\eta _{j}x}.  \label{fxexampj}
\end{equation}
The corresponding Fourier transforms
$F_{j}(k_{\parallel })$ are given by formulae
obtained from Eq. (\ref{Fkexamp}) by the replacements $F_{1}\rightarrow
F_{1}^{(j)}$, $\eta \rightarrow \eta _{j}$. The parameters $F_{1}^{(j)}$ are
defined by the formulae obtained from the corresponding expression for
$F_{1} $ in the paragraph after formula (\ref{Fkexamp}) by the replacements
$\eta \rightarrow \eta _{j}$, $f_{0}\rightarrow f_{0j}$. In figure 3
we have plotted the vacuum pressure on the plate as a function on $\eta a$
for a massless scalar in $D=3$ in the case of
the kernel functions (\ref{fxexampj})
with $\eta _{1}=\eta _{2}\equiv \eta $ and $F_{1}^{(2)}=10$. The numbers
near the curves are the values of the parameter $F_{1}^{(1)}$. For the
values $F_{1}^{(1)}\lesssim -1.08$ the vacuum pressure is negative for all
interplate distances and the corresponding vacuum forces are attractive. For
the values $F_{1}^{(1)}>-1.08$ there are two values of the distance between
the plates for which the vacuum forces vanish. These values correspond to
equilibrium positions of the plates. For the values of the distance in the
region between these positions the vacuum forces acting on plates are
repulsive. Thus, the left equilibrium position is unstable and the right
one is locally stable.

\EPSFIGURE[htp]{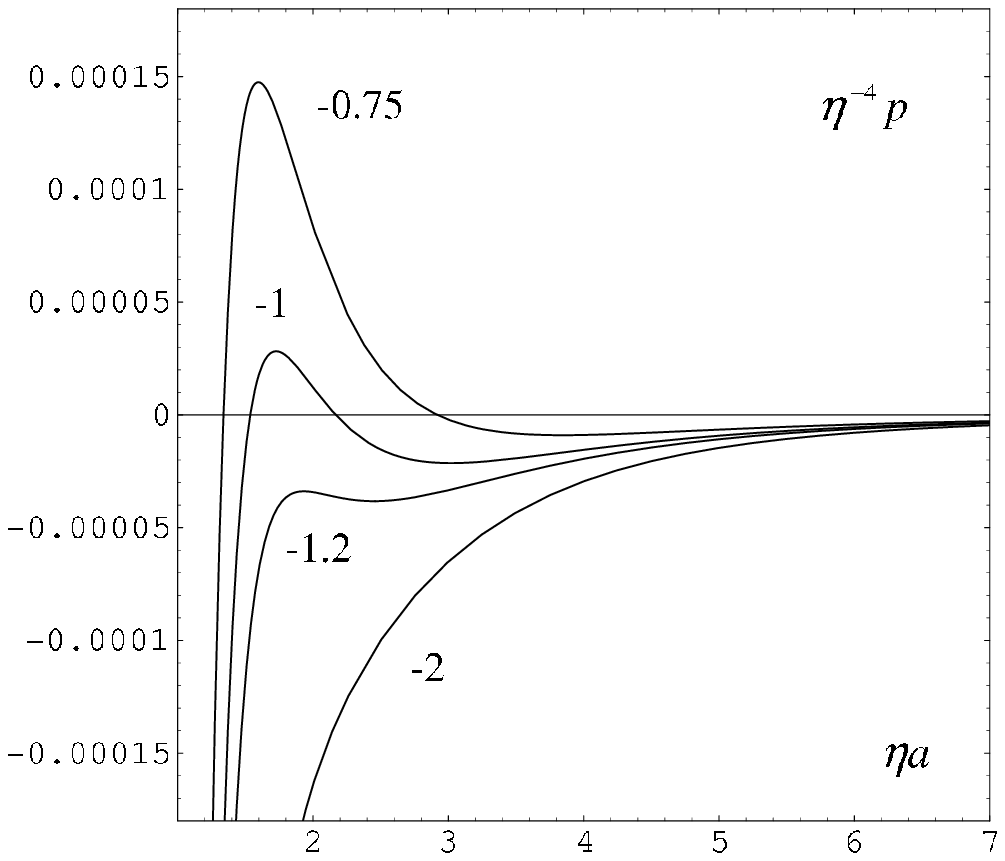,width=.6\textwidth} {Vacuum pressure
on the plate, $p/\protect\eta ^{D+1}$, as a function of
$\protect\eta a$ for a massless scalar in $D=3$ in the case of the
kernel functions (\protect\ref{fxexampj}) with $\protect\eta
_1=\protect\eta _2$ and $F_1^{(2)}=10$. The numbers near the
curves are the values of the parameter $F_1^{(1)}$ defined in the
paragraph after formula (\protect\ref{Fkexamp}).}

\clearpage

\section{Concluding remarks}

\label{sec:conc}

In this paper we have investigated the positive-frequency Wightman function,
the VEVs of the field square and the energy-momentum tensor for a scalar
field with non-local boundary conditions on a single and two parallel plates
in Minkowski spacetime. For the case of a single plate
geometry we have considered the boundary condition of the form (\ref
{Boundcond}), where the kernel function $f(x_{\parallel })$ describes the
properties of the boundary. The VEVs of the physical quantities bilinear in
the field are determined by the Fourier transform
$F(k_{\parallel })$ of this function.
By evaluating the corresponding mode-sum, we have
presented the Wightman function as the sum of a Minkowskian part without
boundary and boundary induced parts, formula (\ref{WF1}). The last term in
this formula corresponds to the contribution of bound states. The
boundary induced part in the VEV of the field square is obtained in the
coincidence limit of the arguments of the Wightman function and is given by
formula (\ref{VEVphi21}). The VEV of the energy-momentum tensor is obtained
by acting with the corresponding second-order differential operator on the
Wightman function and taking the coincidence limit. This VEV is determined
by formula (\ref{VEVemt1}).\ The vacuum stress in the direction
orthogonal to the plate vanishes, and hence the corresponding vacuum
force is zero.

Unlike the case of local boundary conditions, in the
non-local case the vacuum energy-momentum tensor does not vanish for a
conformally coupled massless scalar field. Another important difference is
that for non-local boundary conditions, in general, $\langle
T_{0}^{0}\rangle _{\mathrm{sub}}\neq \langle T_{\mu }^{\mu }\rangle _{
\mathrm{sub}}$, $\mu =2,\ldots ,D$.

As in the case of local boundary
conditions, the energy density and the vacuum stresses diverge on the surface
of the plate. For a nonconformally coupled scalar field the leading term in
the corresponding asymptotic expansion is the same as that for Neumann
boundary condition. As an example, in figures 1 and 2 we have plotted the
vacuum energy density and the vacuum stress as functions of the distance
from the plate for a choice of kernel
function $f$ given by (\ref{fxexamp}).

In section \ref{sec:2pl} we
have considered the geometry of two parallel plates with boundary conditions
(\ref{Boundcond2}). For the region between the plates the corresponding
eigenvalues are solutions of equation (\ref{eigvalues}), where the
coefficients $c_{j}$ are determined by the Fourier transforms of the kernel
functions $f_{j}(x_{\parallel })$ in the boundary conditions. The evaluation
of the corresponding Wightman function is based on a variant of the
generalized Abel--Plana summation formula, Eq. (\ref{sumformula}). The
application of this formula allowed us to extract from the VEVs the parts
resulting from the single plate and to present the part induced from the
second plate in
terms of integrals exponentially convergent for points away from the
boundary. The Wightman function is presented in the form (\ref{WF2pl1}). The
VEVs of the field square and the energy-momentum tensor are obtained from
the Wightman function and are determined by formulae (\ref{phi22pl}), (\ref
{Tmunu2pl}). The vacuum stress in the direction orthogonal to the plates
is uniform. This stress determines the vacuum forces acting on the plates.
The corresponding effective pressure is given by formula (\ref{vacforce}).

For small and large distances between the plates the vacuum forces are
attractive. For intermediate distances the nature of the vacuum forces
depends on the functions in the boundary conditions (\ref{Boundcond2}). For
the example (\ref{fxexampj}) we have shown that,
depending on the parameters, the forces acting on the plates can be
repulsive for intermediate distances (see figure 3). In this case
it is possible to have a locally stable equilibrium value of the
interplate distance stabilized by the vacuum forces.

As far as we know, previous work on non-local boundary conditions for the
Casimir effect had instead considered spectral boundary conditions for spinor
fields in spherically symmetric cavities \cite{Cogn01}, and hence our
results with the non-local boundary conditions (2.2) and (3.1) for scalar
fields are entirely original. From the point of view of bosonic string
theory, a non-local Casimir effect is studied in \cite{Eliz95}, but of a
completely different nature as compared to our work, since one there deals
with a non-local Lagrangian.

\acknowledgments
The work of A. Saharian has been supported by the INFN, by ANSEF Grant No.
05-PS-hepth-89-70, and in part by the Armenian Ministry of Education and
Science, Grant No. 0124. The work of G. Esposito has been partially
supported by PRIN {\it SINTESI}.


\begin{thebibliography}{999}
\bibitem{Espo99a} Esposito G 1999
{\it Class. Quantum Grav.} {\bf 16} 1113

\bibitem{Espo99b} Esposito G 1999
{\it Class. Quantum Grav.} {\bf 16} 3999

\bibitem{Deat91} D'Eath P D and Esposito G 1991
{\it Phys. Rev.} {\bf D 44} 1713

\bibitem{Vasi01} Vassilevich D 2001
{\it JHEP} {\bf 0103} 023

\bibitem{Grub03} Grubb G 2003
{\it Commun. Math. Phys.} {\bf 240} 243

\bibitem{Schr89} Schr\"{o}der M 1989 {\it Rep.
Math. Phys.} {\bf 27} 259

\bibitem{Espo00} Esposito G 2000 {\it Int. J. Mod. Phys.} {\bf A 15} 4539

\bibitem{Rome02} Romeo A and Saharian A A 2002
{\it J. Phys.} {\bf A 35} 1297

\bibitem{Saha04} Saharian A A 2004
{\it Phys. Rev.} {\bf D 69} 085005

\bibitem{Birr82} Birrell N D and Davies P C W
{\it Quantum fields in Curved Space}
(Cambridge University Press, Cambridge, 1982)

\bibitem{Saha05a} Saharian A A 2006 {\it Phys. Rev.} {\bf D 73}
044012

\bibitem{Deut79} Deutsch D and Candelas P 1979 {\it Phys. Rev.}
{\bf D 20} 3063; Kennedy G, Critchley R and Dowker J S 1980 {\it
Ann. Phys. (N.Y.)} {\bf 125} 346; Milton K A 2004 {\it J. Phys.}
{\bf A 37} R209

\bibitem{Ford98} Ford L H and Svaiter N F 1998 {\it Phys. Rev.}
{\bf D 58} 065007

\bibitem{Grah02} Graham N, Jaffe R L, Khemani V, Quandt M,
Scandurra M and Weigel H 2002 {\it Nucl. Phys.} {\bf B 645} 49;
Graham N, Jaffe R L, Khemani V, Quandt M, Scandurra M and Weigel H
2003 {\it Phys. Lett.} {\bf B 572} 196; Graham N and Olum K D 2003
{\it Phys. Rev.} {\bf D 67} 085014

\bibitem{SahaAP} Saharian A A 2000 {\it The generalized Abel--Plana formula.
Applications to Bessel functions and Casimir effect},
Report No. IC/2000/14 [hep-th/0002239]

\bibitem{Saha05b} Saharian A A 2005 {\it Bulk Casimir densities and
vacuum interaction forces in higher dimensional brane models}
[hep-th/0508185].

\bibitem{Cogn01} Cognola G, Elizalde E and Kirsten K 2001 {\it J. Phys.}
{\bf A 34} 7311

\bibitem{Eliz95} Elizalde E and Odintsov S D 1995
{\it Class. Quantum Grav.} {\bf 12} 2881

\end{thebibliography}
\end{document}